\documentclass[a4,11pt]{article}
\usepackage{amssymb}
\usepackage{amsmath}
\usepackage[pdftex]{graphicx}
\usepackage{cite}
\usepackage{slashed}
\usepackage[top=2.5cm, bottom=2.0cm, left=2.5cm, right=2cm]{geometry}

\begin{document}

\begin{center}
{\large \bf
 The effect of the scalar unparticle on the production of Higgs - radion at high energy colliders }\\

\vspace*{1cm}

 {\bf Dang Van Soa$^{a,}$\footnote{soadangvan@gmail.com}, Bui Thi Ha Giang$^{b}$\\}

\vspace*{0.5cm}
$^a$ Hanoi Metropolitian University, 98 Duong Quang Ham, Hanoi, Vietnam\\
 $^b$ Hanoi National University of Education, 136 Xuan Thuy, Hanoi, Vietnam
\end{center}

\begin{abstract}
An attempt is made to present the influence of the scalar unparticle on some scattering processes in the Randall - Sundum model.
The contribution of the scalar unparticle on the production of Higgs - radion at high energy colliders is studied in detail. We evaluate the production cross-sections in the electron-positron ($e^{+}e^{-}$), photon-photon ($\gamma\gamma$) and gluon-gluon ($gg$) collisions, which depend strongly on the collision energy $\sqrt{s}$, the scaling dimension $d_{U}$ of the unparticle operator $\mathcal{O}_{U}$ and the energy scale $\Lambda_{U}$. Numerical evaluation shows that the cross - sections for the pair production of scalar particles are much larger than that of the  associated production of the scalar particle with unparticle under the same conditions.\\

\end{abstract}


\section{Introduction}
The Standard model (SM) is the successful model in describing the elementary particle physics. Recently, the 125 GeV Higgs is discovered by the ATLAS and CMS collaborations \cite{gaa, cha}, which has completed the particle spectrum of the SM. Although the SM has been considered to be successful model, the model suffers from many theoretical drawbacks. In 1999, Lisa Randall and Raman Sundrum suggested the Randall-Sundrum (RS) model to extend the SM and solve the hierarchy problem naturally \cite{rs,frank}. The RS setup involves two three-branes bounding a slice of 5D compact anti-de Sitter space. Gravity is localized at the UV brane, while the SM fields are supposed to be localized at the IR brane. The separation between the two 3-branes leads directly to the existence of an additional scalar called the radion ($\phi$ ), corresponding to the quantum fluctuations of the distance between the two 3-branes \cite{dominici,csa,gold}. \\
\hspace*{0.5cm}In the Lagrangian of the Standard model, the scale invariance is broken at or above the electroweak scale \cite{zhang, cheung}. The scale invariant sector has been considered as an effective theory at TeV scale and that if it exists, it is made of unparticle suggested by Geogri \cite{georgi,georgi2}. Based on the Banks-Zaks theory \cite{banks}, unparticle stuff with nontrivial scaling dimension is considered to exist in our world. The invariant Banks-Zaks field can be connected to the SM particles \cite{chenhe}. Recently, the possibility of the unparticle has been studied with CMS detector at the LHC \cite{cms15, cms16}.  \\
\hspace*{0.5cm}The effects of unparticle on properties of high energy colliders have been intensively studied in Refs.\cite{pra,alan,maj, kuma,sahi,kiku,chen,kha,alie, fried,anto}. However, the influence of scalar unparticle on the production of particles at the high energy colliders have not yet been concerned in the RS model. In this work, the contribution of the scalar unparticle on  the production of Higgs - radion  at the $e^{+}e^{-}$, $\gamma\gamma$ and $gg$ colliders are studied in detail. 
The layout of this paper is as follows. In Section II, we give a review of the RS model and the mixing of Higgs-radion.
 The contribution of the scalar unparticle on the production of Higgs - radion at high energy colliders are calculated in Section III. Finally, we summarize our results and make conclusions in Section IV.

\section{A review of Randall-Sundrum model and the mixing of Higgs-radion}
The RS model is based on a 5D spacetime with non - factorizable geometry. The single extra dimension is compactified on an $S^{1} /Z_{2}$ orbifold of which two fixed points accommodate two three-branes (4D hyper-surfaces), the UV brane and the IR brane. The four dimensional effective action is obtained by integrating out the extra dimension. The classical action describing the above set-up is given by \cite{rs}
\begin{equation}
S = S_{gravity} + S_{IR} + S_{UV},
\end{equation}
\begin{subequations}
\begin{align}
S_{gravity}=\int d^5 x \sqrt{-G}\left(-\Lambda + 2M^3 R\right),\\
S_{IR}=\int d^4 x \sqrt{-g_{IR}}(\mathcal{L}_{IR}-V_{IR}),\\
S_{UV}=\int d^4 x \sqrt{-g_{UV}}(\mathcal{L}_{UV}-V_{UV}), 
\end{align}
\end{subequations}
where M is the five dimensional Planck scale, $G = detG_{MN}$, $\Lambda$ is a bulk cosmological constant, R is the 5D Ricci scalar. In the RS model, the values of the bare parameters are determined by the Planck scale and the applicable value for size of the extra dimension is assessed by $kr_{c} \pi \simeq 35$ ($r_{c}$ - the compactification radius and $k$ - the bulk curvature). Thus the weak and the gravity scales can be naturally generated. Consequently, the hierarchy problem is addressed. The gravity-scalar mixing is described by the following action\cite{dominici}
\begin{equation}
S_{\xi } =\xi \int d^{4}x \sqrt{g_{vis} } R(g_{vis} )\hat{H}^{+} \hat{H},
\end{equation}
where $\xi $ is the mixing parameter, $R(g_{vis})$ is the Ricci scalar for the metric $g_{vis}^{\mu \nu } =\Omega _{b}^{2} (x)(\eta ^{\mu \nu } +\varepsilon h^{\mu \nu } )$ induced on the visible brane, $\Omega _{b} (x) = e^{-kr_{c} \pi} (1 + \frac{\phi_{0}}{\Lambda _{\phi }})$ is the warp factor, $\phi_{0}$ is the canonically normalized massless radion field, $\hat{H}$ is the Higgs field in the 5D context before rescaling to canonical normalization on the brane.
 With $\xi \ne 0$, there is neither a pure Higgs boson nor pure radion mass eigenstate. This $\xi$ term mixes the $h_{0}$ and $\phi_{0}$ into the mass eigenstates $h$ and $\phi$ as given by 
\begin{equation} 
\left(\begin{array}{c} {h_{0} } \\ {\phi _{0} } \end{array}\right)=\left(\begin{array}
{cc} {1} & {6\xi \gamma /Z} \\ {0} & {-1/Z} \end{array}\right)\left(\begin{array}{cc}
 {\cos \theta } & {\sin \theta } \\ {-\sin \theta } & {\cos \theta } \end{array}\right)
 \left(\begin{array}{c} {h} \\ {\phi } \end{array}\right)=\left(\begin{array}{cc}
  {d} & {c} \\ {b} & {a} \end{array}\right)\left(\begin{array}{c} {h} \\ {\phi } \end{array}\right), \label{pt1}
\end{equation}
where
$Z^{2} = 1 + 6\gamma ^{2} \xi \left(1 -\, \, 6\xi \right) = \beta - 36\xi ^{2}\gamma ^{2}$ is the coefficient of the radion kinetic term after undoing the kinetic mixing, $\gamma = \upsilon /\Lambda _{\phi }, \upsilon = 246$ GeV, $a = -\dfrac{cos\theta}{Z}, b = \dfrac{sin\theta}{Z}, c = sin\theta + \dfrac{6\xi\gamma}{Z}cos\theta, d = cos\theta - \dfrac{6\xi\gamma}{Z}sin\theta$. The mixing angle $\theta $ is
\begin{equation}
\tan 2{\theta } = 12{\gamma \xi Z}\frac{m_{h_{0}}^{2}}{m_{\phi _{0}}^{2} - m_{h_{0}}^{2} \left( Z^{2} - 36\xi^{2} \gamma ^{2} \right)},
\end{equation}
where $m_{h_{0}}$ and $m_{\phi _{0}}$ are the Higgs and radion masses before mixing. The new physical fields h and $\phi $ in (\ref{pt1}) are Higgs-dominated state and radion, respectively
\begin{equation} 
m_{h,\phi }^{2} =\frac{1}{2Z^{2} } \left[m_{\phi _{0} }^{2} +\beta m_{h_{0} }^{2} \pm \sqrt{(m_{\phi _{0} }^{2} +\beta m_{h_{0} }^{2} )^{2} -4Z^{2} m_{\phi _{0} }^{2} m_{h_{0} }^{2} } \right].
\end{equation}
\\
Feynman rules for the couplings of Higgs and radion are showed as follows
\begin{align}
&V(h, f, \overline{f}) = i\overline{g}_{f\overline{f}h} = -i\dfrac{gm_{f}}{2m_{W}}\left( d + \gamma b\right),\\
&V(\phi, f, \overline{f}) = i\overline{g}_{f\overline{f}\phi} = -i\dfrac{gm_{f}}{2m_{W}}\left( c + \gamma a\right),\\
&\begin{aligned}
V(h, \gamma_{\mu}(k_{1}), \gamma_{\mu}(k_{2})) = &iC_{\gamma h}\left[(k_{1}k_{2})\eta^{\mu\nu} - k_{1}^{\nu}k_{2}^{\mu}\right]\\
= &-i\dfrac{\alpha}{2\pi\upsilon_{0}}\left((d + \gamma b)\sum_{i}e_{i}^{2}N_{c}^{i}F_{i}(\tau_{i}) - (b_{2} + b_{Y})\gamma b \right)\\
&\times \left[(k_{1}k_{2})\eta^{\mu\nu} - k_{1}^{\nu}k_{2}^{\mu}\right],
\end{aligned}
\end{align}
\begin{equation}
\begin{aligned}
V(\phi, \gamma_{\mu}(k_{1}), \gamma_{\mu}(k_{2})) =
&iC_{\gamma \phi}\left[(k_{1}k_{2})\eta^{\mu\nu} - k_{1}^{\nu}k_{2}^{\mu}\right]\\
= & -i\dfrac{\alpha}{2\pi\upsilon_{0}}\left((c + \gamma a)\sum_{i}e_{i}^{2}N_{c}^{i}F_{i}(\tau_{i}) - (b_{2} + b_{Y})\gamma a \right)\\
&\times \left[(k_{1}k_{2})\eta^{\mu\nu} - k_{1}^{\nu}k_{2}^{\mu}\right].
\end{aligned}
\end{equation}
\begin{align}
&\begin{aligned}
V(h, g_{\mu, a}(k_{1}), g_{\mu, b}(k_{2})) = &iC_{g h}\delta^{ab}\left[(k_{1}k_{2})\eta^{\mu\nu} - k_{1}^{\nu}k_{2}^{\mu}\right]\\
= &-i\dfrac{\alpha_{s}}{4\pi\upsilon_{0}}\left((d + \gamma b)\sum_{i}F_{1/2}(\tau_{i}) - 2b_{3}\gamma b \right)\\
&\times \delta^{ab}\left[(k_{1}k_{2})\eta^{\mu\nu} - k_{1}^{\nu}k_{2}^{\mu}\right],
\end{aligned}\\
&\begin{aligned}
V(\phi, g_{\mu, a}(k_{1}), g_{\mu, b}(k_{2})) = &iC_{g\phi}\delta^{ab}\left[(k_{1}k_{2})\eta^{\mu\nu} - k_{1}^{\nu}k_{2}^{\mu}\right]\\
= &-i\dfrac{\alpha_{s}}{4\pi\upsilon_{0}}\left((c + \gamma a)\sum_{i}F_{1/2}(\tau_{i}) - 2b_{3}\gamma a \right)\\
&\times \delta^{ab}\left[(k_{1}k_{2})\eta^{\mu\nu} - k_{1}^{\nu}k_{2}^{\mu}\right],
\end{aligned}
\end{align}
where $b_{3} = 7, b_{2}= 19/6, b_{Y} = - 41/6$ are the $\mathrm{SU}(2)_{L} \otimes
 \mathrm{U}(1)_{Y}$ $\beta$-function coefficients in the SM.\\
The auxiliary functions of the $h$ and $\phi$ are given by
\begin{align}
&F_{1/2}(\tau) = -2 \tau[1 + (1-\tau) f(\tau)],\\
&F_1(\tau) = 2 + 3\tau + 3\tau(2-\tau) f(\tau),
\end{align}
with
\begin{align}
&f(\tau ) = \left(\sin ^{-1} \frac{1}{\sqrt{\tau } } \right)^{2} \, \, \, (for\, \, \, \tau >1),\\
&f(\tau ) = -\frac{1}{4} \left(\ln \frac{\eta _{+} }{\eta _{-} } - i\pi \right)^{2} \, \, \, (for\, \, \, \tau <1),\\
&\eta _{\pm} = 1\pm \sqrt{1 - \tau } ,\, \, \, \tau _{i} = \left(\frac{2m_{i} }{m_{s} } \right)^{2} .
\end{align}
 $m_{i}$ is the mass of the internal loop particle (including quarks, leptons and W boson), $m_{s}$  is the mass of the scalar state ($h$ or $\phi$). Here, $\tau _{f} = \left(\frac{2m_{f} }{m_{s} } \right)^{2},   \tau _{W} = \left(\frac{2m_{W} }{m_{s} } \right)^{2}$ denote the squares of fermion and W gauge boson mass ratios, respectively. There are four independent parameters $\Lambda _{\phi } ,\, \, m_{h} ,\, \, m_{\phi } ,\, \, \xi$ that must be specified to fix the state mixing parameters. We consider the case of $\Lambda _{\phi } = 5$ TeV and $\frac{m_{0} }{M_{P} } = 0.1$, which makes the radion stabilization model most natural \cite{csa,gold}.
\section{ The contribution of the scalar unparticle on the production of Higgs - radion at high energy colliders}
The effects of unparticle on properties of high energy colliders have been intensively studied in Refs.\cite{pra,alan,maj, kuma,sahi,kiku,chen,kha,alie, fried,anto}. 
In the rest of this work, we restrict ourselves by considering only scalar unparticle. The scalar unparticle propagator is given by \cite{cheung,georgi2}
\begin{equation}
\Delta_{scalar} = \dfrac{iA_{d_{U}}}{2sin(d_{U}\pi)}(-q^{2})^{d_{U}-2},
\end{equation}
where 
\begin{align}
&A_{d_{U}} = \dfrac{16\pi^{2}\sqrt{\pi}}{(2\pi)^{2d_{U}}}\dfrac{\Gamma\left( d_{U} + \dfrac{1}{2}\right) }{\Gamma(d_{U} - 1)\Gamma (2d_{U})},\\
&(-q^{2})^{d_{U}-2} = \begin{cases}
|q^{2}|^{d_{U} - 2} e^{-d_{U}\pi}   \text{   for s-channel process}, q^{2}  \text{ is positive,}\\
|q^{2}|^{d_{U} - 2}                \text{   for u-, t-channel process}, q^{2}  \text{ is negative.}
\end{cases}
\end{align}
\hspace*{1cm}The effective interactions for the scalar unparticle operators are given by
\begin{align}
&\lambda_{ff}\dfrac{1}{\Lambda^{d_{U}-1}_{U}}\overline{f}f O_{U}, \lambda_{ff}\dfrac{1}{\Lambda^{d_{U}-1}_{U}}\overline{f} i\gamma^{5}f O_{U}, \lambda_{ff}\dfrac{1}{\Lambda^{d_{U}}_{U}}\overline{f} \gamma^{\mu}f (\partial_{\mu}O_{U}), \lambda_{gg}\dfrac{1}{\Lambda^{d_{U}}_{U}}G_{\alpha\beta}G^{\alpha\beta} O_{U},
&\lambda_{hh}\dfrac{1}{\Lambda^{d_{U}-2}_{U}}H^{+}H O_{U},
\end{align}
where $G^{\alpha\beta}$ denotes the gauge field strength and $f$ stands for a standard model fermion.
Feynman rules for the couplings of the scalar unparticle in the RS model are showed as follows
\begin{align}
&g_{f\overline{f}U} = i\overline{g}_{f\overline{f}U} = i\dfrac{\lambda_{ff}}{\Lambda_{U}^{d_{U} - 1}},\\
&g_{\gamma\gamma U} = -i\overline{g}_{\gamma\gamma U}\left[(p_{1}p_{2})\eta^{\mu\nu} - p_{1}^{\nu}p_{2}^{\mu}\right] = -4i\dfrac{\lambda_{\gamma\gamma}}{\Lambda_{U}^{d_{U}}} \left[(p_{1}p_{2})\eta^{\mu\nu} - p_{1}^{\nu}p_{2}^{\mu}\right] ,\\
&g_{ggU} = - i\overline{g}_{ggU}\left[(p_{1}p_{2})\eta^{\mu\nu} - p_{1}^{\nu}p_{2}^{\mu}\right]  = -4i\dfrac{\lambda_{gg}}{\Lambda_{U}^{d_{U}}} \left[(p_{1}p_{2})\eta^{\mu\nu} - p_{1}^{\nu}p_{2}^{\mu}\right] , \\
&g_{hhU} = -i\overline{g}_{hhU} = -i\dfrac{\lambda_{hh}}{\Lambda_{U}^{d_{U} - 2}},\\
&g_{\phi\phi U} = - i\overline{g}_{\phi\phi U} = -i\dfrac{\lambda_{\phi\phi}}{\Lambda_{U}^{d_{U} - 2}}.  
\end{align}
\hspace*{0.5cm}Using the above formulas, we will study the effect of the scalar unparticle on some high energy scatterings in the RS model. We note here that in our previours works \cite{soa,soa1,soael} we have shown that the detection of scalar particles in the RS model at high energy colliders would provide a clear evidence of new physics beyond the SM. Now we will investigate the contribution of the scalar unparticle on the production of Higgs - radion in the RS model at high energy colliders, such as $e^{+}e^{-}$, $\gamma\gamma$ and $g g$ collisions in which Feynman diagrams are considered in detail in Appendix A.\\
\textbf{1. The $e^{+}e^{-}$ $\rightarrow$ $hh/\phi\phi$ collisions}\\
Now we consider the collision process in which the initial state contains electron and positron, the final state contains the couple of the scalar particles (Higgs or radion). We note here that the contribution of the scalar unparticle is by the propagator in the s - channel
\begin{equation} \label{pt4}
e^{-}(p_{1}) + e^{+}(p_{2}) \    \underrightarrow{\phi, h, U} \         X (k_{1}) + X (k_{2}),
\end{equation}
where $X$ is Higgs or radion. The transition amplitude is given by
\begin{align}
&\begin{aligned}
M_{s} = &-i\left( \dfrac{\overline{g}_{ee\phi}\overline{g}_{XX\phi}}{q_{s}^{2} - m^{2}_{\phi}} + \dfrac{\overline{g}_{eeh}\overline{g}_{XXh}}{q_{s}^{2} - m^{2}_{h}} \right)\overline{v}(p_{2})u(p_{1})\\
& -i \overline{g}_{eeU}\overline{g}_{XXU} \dfrac{A_{d_{U}}}{2sin(d_{U}\pi)} (-q_{s}^{2})^{d_{U} - 2}\overline{v}(p_{2})u(p_{1}) ,
\end{aligned}\\
&M_{u} = -i\overline{g}_{eeX}\overline{g}_{eeX}\overline{v}(p_{2})\dfrac{(\slashed{q}_{u} + m_{e})}{q_{u}^{2} - m^{2}_{e}}u(p_{1}),\\
&M_{t} = -i\overline{g}_{eeX}\overline{g}_{eeX}\overline{v}(p_{2})\dfrac{(\slashed{q}_{t} + m_{e})}{q_{t}^{2} - m^{2}_{e}}u(p_{1}).
\end{align}
Here, $q_{s} = p_{1} + p_{2} = k_{1} + k_{2},  q_{u} = p_{1} - k_{2} = k_{1} - p_{2}, q_{t} = p_{1} - k_{1} = k_{2} - p_{2}$ and $s = (p_{1} + p_{2})^{2}$ is the square of the collision energy. From the expressions of the differential cross-section \cite{pes}
\begin{equation}
\frac{d\sigma}{d(cos\psi)} = \frac{1}{32 \pi s} \frac{|\overrightarrow{k}_{1}|}{|\overrightarrow{p}_{1}|} |M_{fi}|^{2},
\end{equation}
where $|M_{fi}|^{2} = |M_{s}|^{2} + |M_{u}|^{2} + |M_{t}|^{2} + Re\left( M_{s}^{*}M_{u} + M_{u}^{*}M_{s}+ M_{s}^{*}M_{t} + M_{t}^{*}M_{s} + M_{u}^{*}M_{t} + M_{t}^{*}M_{u}\right)$, $\psi = (\overrightarrow{p}_{1}, \overrightarrow{k}_{1})$ is the scattering angle. The model parameters are chosen as: $\lambda_{ff} = \lambda_{hh} = \lambda_{\phi\phi} = \lambda_{0} = 1$, $\Lambda_{U} = 1000$ GeV, $1 < d_{U} < 2$ in case of the scalar unparticle \cite{fried}, $ m_{h}$ = 125 GeV, $m_{\phi}$ = 10 GeV\cite{soa,soa1}. We give estimates for the cross-sections which depend on the collision energy $\sqrt{s}$, the scaling dimension $d_{U}$ of the unparticle operator $\mathcal{O}_{U}$ and the energy scale $\Lambda_{U}$ as follows \\
i) In Fig.1 we plot the total cross-sections as the function of $d_{U}$. The collision energy is chosen as $\sqrt{s} = 500$ GeV and $1.1 \leq d_{U} \leq 1.9$. From the Fig.1 we can see that {\it in case of the additional scalar unparticle propagator}, the cross sections decrease rapidly as $d_{U}$ increases and they are flat when $d_{U} > 1.6 $.\\
ii) In Fig.2 we evaluate the dependence of the total cross-sections on the collision energy $\sqrt{s}$. The collision energy is chosen in the range of 500 GeV$ \leq \sqrt{s} \leq$ 1000 GeV (ILC), the various $d_{U}$ is chosen as 1.1, 1.3, 1.5, 1.7, respectively. The figure shows that the total cross-sections decrease when the collision energy $\sqrt{s}$ increases.
{\it It is worth noting that with the contribution of the scalar unparticle propagator, the cross-sections for pair production of scalar particles are much enhanced}. \\
iii) In Fig.3 we evaluate the dependence of the total cross-sections on the $\Lambda_{U}$ at the fixed collision energy, $\sqrt{s} = 500$ GeV. In case of the additional scalar unparticle propagator, the cross-sections increase rapidly in the region of 2 TeV $\leq \Lambda_{U} \leq$ 5 TeV. Note that here we only plot the maximum cross-sections based on Fig.1.\\
\textbf{2. The $e^{+}e^{-} \rightarrow Uh/U\phi$ collisions}\\
In this section, we investigate the associated production of the scalar particle with unparticle at high energy $e^{+}e^{-}$ colliders in which  the scalar unparticle contribution on the scattering process is in the final state 
\begin{equation}
e^{-}(p_{1}) + e^{+}(p_{2})     \underrightarrow{\phi, h} \         X (k_{1}) + U (k_{2}).
\end{equation}
The transition amplitude can be given as follows
\begin{align}
&M_{s} = i \dfrac{\overline{g}_{eeX}\overline{g}_{XXU}}{q_{s}^{2} - m^{2}_{X}}\overline{v}(p_{2})u(p_{1})\\
&M_{u} = - i\dfrac{\overline{g}_{eeX}\overline{g}_{eeU}}{q_{u}^{2} - m^{2}_{e}}\overline{v}(p_{2})(\slashed{q}_{u} + m_{e}) u(p_{1}),\\
&M_{t} = - i\dfrac{\overline{g}_{eeX}\overline{g}_{eeU}}{q_{t}^{2} - m^{2}_{e}}u(p_{1})(\slashed{q}_{t} + m_{e})\overline{v}(p_{2}) .
\end{align}
With the parameters chosen as above, we give some estimates for the cross-sections with the contribution of scalar unparticle as follows \\
i) In Fig.4 we plot the total cross-sections as the function of $d_{U}$. We can see from the figure that the curve of the cross-sections is similar to Fig.1. That is, the cross-sections decrease rapidly as $d_{U}$ increases.  \\
ii) In Fig.5 we evaluate the dependence of the total cross-sections on the collision energy $\sqrt{s}$ with the various $d_{U}$.
The result shows that the cross-sections decrease as the collision energy $\sqrt{s}$ increases. Note that the curve of the cross-sections is flat at very high energies. \\ 
iii) The dependence of the total cross-sections on the $\Lambda_{U}$ at the fixed collision energy, $\sqrt{s} = 500$ GeV is shown Fig.6. The figures show that the total cross-section for the associated production in the $e^{+} e^{-} \rightarrow Uh$ collision is about $10^{3}$ times larger than that in the $e^{+} e^{-}\rightarrow U\phi$ collision. Numerical values for the production cross section  with $d_{U} = 1.1$ are given in detail in Table 1. We can see from Table 1 that the cross-sections for the pair production of scalar particles are much larger than that of the  associated production of scalar particles with unparticle under the same conditions. It is worthing to note that, when the collision energy increases, the total cross-section in the $e^{+} e^{-} \rightarrow \phi\phi$ collision is insignificantly larger than that in $e^{+} e^{-} \rightarrow hh$ collision.\\  
\textbf{3. The $\gamma\gamma \rightarrow hh/\phi\phi $ collisions}\\
In this section, we consider the collision process in which the initial state contains the couple of photons, the final state contains the couple of scalar particles. The Feynman diagram is given by
 \begin{equation}
\gamma(p_{1}) + \gamma(p_{2})     \underrightarrow{\phi, h, U} \         X (k_{1}) + X (k_{2}).
\end{equation}
We obtain the results in the $s, u, t$ - channels
\begin{align}
&\begin{aligned}
 M_{s} = &\left[- \left(C_{\gamma\phi}\frac{\overline{g}_{\phi XX}}{q^{2}_{s} - m_{\phi}^{2}} + C_{\gamma h}\frac{\overline{g}_{hXX}}{q^{2}_{s} - m_{h}^{2}} \right)- 4i\overline{g}_{XXU}\dfrac{\lambda_{\gamma\gamma}}{\Lambda_{U}^{d_{U}}}\dfrac{A_{d_{U}}}{2sin(d_{U}\pi)} (-q_{s}^{2})^{d_{U} - 2} \right]\\
 &\times \varepsilon_{\alpha} (p_{1}) \left[\left(p_{1}p_{2}\right)\eta^{\alpha\beta} - p_{1}^{\beta}p_{2}^{\alpha}\right]\varepsilon_{\beta} (p_{2}),
\end{aligned}\\
&M_{u} = iC_{\gamma X}C_{\gamma X}\frac{1}{q_{u}^{2}}\left[\left(p_{2}q_{u}\right)\eta^{\nu}_{\rho} - p_{2\rho}q_{u}^{\nu}\right]\varepsilon_{\nu}(p_{2})\left[\left(p_{1}q_{u}\right)\eta^{\mu\rho} - p_{1}^{\rho}q_{u}^{\mu}\right]\varepsilon_{\mu} (p_{1}),\\
&M_{t} = iC_{\gamma X}C_{\gamma X}\frac{1}{q_{t}^{2}}\left[\left(p_{1}q_{t}\right)\eta^{\mu}_{\sigma} - p_{1\sigma}q_{t}^{\mu}\right]\varepsilon_{\mu}(p_{1})\left[\left(p_{2}q_{t}\right)\eta^{\nu\sigma} - p_{2}^{\sigma}q_{t}^{\nu}\right]\varepsilon_{\nu} (p_{2}).
\end{align}
Now we estimate the production cross-sections with the contribution of the scalar unparticle propagator as follows \\
i) In Fig.7 we plot the total cross-sections in the $\gamma\gamma \rightarrow hh/\phi\phi $ collisions as the function of $d_{U}$. The collision energy is chosen as $\sqrt{s} = 3000$ GeV (CLIC) and $1.1 \leq d_{U} \leq 1.9$. We can see from the figure that, the curve goes through the minimum value at $d_{U}= 1.65 $ and then increases rapidly with $d_{U}$.\\
ii) In Fig.8 we plot the total cross-sections as a function of the collision energy $\sqrt{s}$. The collision energy region is $1  TeV \leq \sqrt{s} \leq 5 TeV$. The total cross-sections decrease gradually as $\sqrt{s}$ increases with the fixed $d_{U}$.\\
iii) In Fig.9 we plot the dependence of the total cross-sections on the energy scale $\Lambda_{U}$ with the parameters chosen as above. The figure shows that the cross-sections decrease gradually as the $\Lambda_{U}$ increases. \\
\textbf{4. The $\gamma\gamma \rightarrow Uh/U\phi $ collisions}\\
In this section, we investigate the unparticle contribution on $\gamma\gamma \rightarrow Uh/U\phi$ collisions
 \begin{equation}
\gamma(p_{1}) + \gamma(p_{2})     \underrightarrow{\phi, h} \         X (k_{1}) + U (k_{2}).
\end{equation}
The transition amplitude can be written as follows
\begin{align}
& M_{s} = iC_{\gamma X}\frac{\overline{g}_{XXU}}{q^{2}_{s} - m_{X}^{2}} \varepsilon_{\alpha} (p_{1}) \left[\left(p_{1}p_{2}\right)\eta^{\alpha\beta} - p_{1}^{\beta}p_{2}^{\alpha}\right]\varepsilon_{\beta} (p_{2}),\\
&M_{u} = -iC_{\gamma X}\overline{g}_{\gamma\gamma U}\frac{1}{q_{u}^{2}}\left[\left(p_{2}q_{u}\right)\eta^{\nu}_{\rho} - p_{2\rho}q_{u}^{\nu}\right]\varepsilon_{\nu}(p_{2})\left[\left(p_{1}q_{u}\right)\eta^{\mu\rho} - p_{1}^{\rho}q_{u}^{\mu}\right]\varepsilon_{\mu} (p_{1}),\\
&M_{t} = -iC_{\gamma X}\overline{g}_{\gamma\gamma U}\frac{1}{q_{t}^{2}}\left[\left(p_{1}q_{t}\right)\eta^{\mu}_{\sigma} - p_{1\sigma}q_{t}^{\mu}\right]\varepsilon_{\mu}(p_{1})\left[\left(p_{2}q_{t}\right)\eta^{\nu\sigma} - p_{2}^{\sigma}q_{t}^{\nu}\right]\varepsilon_{\nu} (p_{2}).
\end{align}
We estimate the cross-sections for the associated production as follows \\
i) In Fig.10 we plot the total cross-sections as the function of $d_{U}$ with the parameters chosen as in previous items.
 The figure shows that the curve of the cross section is similar to Fig.1. We can see that the cross section decreases rapidly as $d_{U}$ increases and it is flat with $d_{U} > 1.6 $. \\
ii) In Fig.11 we evaluate the dependence of the total cross-sections on the collision energy $\sqrt{s}$ with the fixed $d_{U}$. The figure shows that when the collision energy $\sqrt{s}$ increases then the total cross-sections increase gradually. \\
iii) In Fig.12 we plot the dependence of the total cross-sections on the $\Lambda_{U}$. The figure shows that in the region 1 TeV $\leq \Lambda_{U} \leq$ 5 TeV the cross-sections decrease gradually as $\Lambda_{U}$ increases.\\
Some typical values for cross-sections are given in detail in Table 2. The result shows that the cross-sections for pair production of scalar particles are much larger than that of the associated production. Moreover, the total cross-section in $\gamma\gamma \rightarrow Uh$ collision is larger than that in $\gamma\gamma \rightarrow U\phi$ collision
under the same conditions. \\
\textbf{5. The $gg \rightarrow hh/\phi\phi $ collisions}\\
Now we consider the $gg \rightarrow hh/\phi\phi $ process which is similar to the $\gamma\gamma \rightarrow hh/\phi\phi $ process. The reaction is given by
 \begin{equation} 
g(p_{1}) + g(p_{2})     \underrightarrow{\phi, h, U} \         X (k_{1}) + X (k_{2}).
\end{equation}
The transition amplitude for this process can be written as
\begin{align}
&\begin{aligned}
 M_{s} = &\left[- \left(C_{g\phi}\frac{\overline{g}_{\phi XX}}{q^{2}_{s} - m_{\phi}^{2}} + C_{gh}\frac{\overline{g}_{hXX}}{q^{2}_{s} - m_{h}^{2}} \right)- 4i\overline{g}_{XXU}\dfrac{\lambda_{gg}}{\Lambda_{U}^{d_{U}}}\dfrac{A_{d_{U}}}{2sin(d_{U}\pi)} (-q_{s}^{2})^{d_{U} - 2} \right]\\
 &\times \varepsilon_{\alpha} (p_{1}) \left[\left(p_{1}p_{2}\right)\eta^{\alpha\beta} - p_{1}^{\beta}p_{2}^{\alpha}\right]\varepsilon_{\beta} (p_{2}),
\end{aligned}\\
&M_{u} = iC_{gX}C_{gX}\frac{1}{q_{u}^{2}}\left[\left(p_{2}q_{u}\right)\eta^{\nu}_{\rho} - p_{2\rho}q_{u}^{\nu}\right]\varepsilon_{\nu}(p_{2})\left[\left(p_{1}q_{u}\right)\eta^{\mu\rho} - p_{1}^{\rho}q_{u}^{\mu}\right]\varepsilon_{\mu} (p_{1}),\\
&M_{t} = iC_{gX}C_{gX}\frac{1}{q_{t}^{2}}\left[\left(p_{1}q_{t}\right)\eta^{\mu}_{\sigma} - p_{1\sigma}q_{t}^{\mu}\right]\varepsilon_{\mu}(p_{1})\left[\left(p_{2}q_{t}\right)\eta^{\nu\sigma} - p_{2}^{\sigma}q_{t}^{\nu}\right]\varepsilon_{\nu} (p_{2}).
\end{align}
We evaluate the cross-sections as follows \\
i) In Fig.13 we plot the total cross-sections in the $gg \rightarrow hh/\phi\phi $ collisions as the function of $d_{U}$. We can see from the figure that the shape of the cross-section is similar to Fig.7. That is, 
 the curve of the cross-sections goes through the minimum value and then increases rapidly with $d_{U}$.\\
ii) In Fig.14 we plot the total cross-sections as a function of the collision energy $\sqrt{s}$. The figure shows that, the cross-sections decrease as $\sqrt{s}$ increases. The total cross-section in $gg \rightarrow \phi\phi$ collision is insignificantly larger than that in $gg \rightarrow  hh$ collision.\\
iii)  In Fig.15 we plot the dependence of the total cross-sections on the $\Lambda_{U}$. We can see that in the region 1 TeV $\leq \Lambda_{U} \leq$ 5 TeV, the cross-sections decrease as $\Lambda_{U}$ increases.\\
\textbf{6. The $gg \rightarrow Uh/U\phi $ collisions}\\
Finally, we study the contribution of the scalar unparticle on the associated production in $gg \rightarrow Uh/U\phi$ collisions 
 \begin{equation}
g(p_{1}) + g(p_{2})     \underrightarrow{\phi, h} \         X (k_{1}) + U (k_{2}).
\end{equation}
We obtain the transition amplitude in the $s, u, t$ - channels
\begin{align}
& M_{s} = iC_{g X}\frac{\overline{g}_{XXU}}{q^{2}_{s} - m_{X}^{2}} \varepsilon_{\alpha} (p_{1}) \left[\left(p_{1}p_{2}\right)\eta^{\alpha\beta} - p_{1}^{\beta}p_{2}^{\alpha}\right]\varepsilon_{\beta} (p_{2}),\\
&M_{u} = -iC_{g X}\overline{g}_{ggU}\frac{1}{q_{u}^{2}}\left[\left(p_{2}q_{u}\right)\eta^{\nu}_{\rho} - p_{2\rho}q_{u}^{\nu}\right]\varepsilon_{\nu}(p_{2})\left[\left(p_{1}q_{u}\right)\eta^{\mu\rho} - p_{1}^{\rho}q_{u}^{\mu}\right]\varepsilon_{\mu} (p_{1}),\\
&M_{t} = -iC_{g X}\overline{g}_{ggU}\frac{1}{q_{t}^{2}}\left[\left(p_{1}q_{t}\right)\eta^{\mu}_{\sigma} - p_{1\sigma}q_{t}^{\mu}\right]\varepsilon_{\mu}(p_{1})\left[\left(p_{2}q_{t}\right)\eta^{\nu\sigma} - p_{2}^{\sigma}q_{t}^{\nu}\right]\varepsilon_{\nu} (p_{2}).
\end{align}
We estimate the cross-sections for associated production as follows \\
i) In Fig.16 we plot the total cross-sections as the function of $d_{U}$. From the figure we can see that the cross section decreases rapidly as $d_{U}$ increases and it is flat when  $d_{U} > 1.45 $.\\
ii) In Fig.17 we evaluate the dependence of the total cross-sections on the collision energy $\sqrt{s}$ with the fixed $d_{U}$. The figure shows that when the collision energy $\sqrt{s}$ increases in the region $1 TeV \leq \sqrt{s} \leq 5 TeV$ then the total cross-sections increase. The total cross-section in $gg \rightarrow Uh$ collision is larger than that in $gg \rightarrow U\phi$ collision. \\
iii) In Fig.18 we plot the dependence of the total cross-sections on the $\Lambda_{U}$. The figure shows that the cross-sections decrease as $\Lambda_{U}$ increases. Some numerical values for cross sections in case of $d_{U} = 1.1$ are given in Table 3.
\section{Conclusion}
In this paper, we have evaluated the contribution of the scalar unparticle on the production cross-sections of Higgs - radion in the Randall- Sundrum model at the ($e^{+}e^{-}$), ($\gamma\gamma$) and ($gg$) colliders, which depend strongly on the collision energy $\sqrt{s}$, the scaling dimension $d_{U}$ of the unparticle operator $\mathcal{O}_{U}$ and the energy scale $\Lambda_{U}$. The results indicate that the cross - sections for the pair production of scalar particles are much larger than that of the  associated production of scalar particle with the unparticle under the same conditions.\\
\hspace*{0.5cm}In the $e^{+}e^{-}\rightarrow hh /\phi\phi$ collisions, the production cross - section decreases as the collision energy $\sqrt{s}$ increases. With the contribution of the scalar unparticle propagator, the cross-sections for the pair production of scalar particles are much enhanced while the cross-sections for the associated production in $e^{+} e^{-} \rightarrow  Uh/U\phi $ collisions are very small. Numerical evaluation has shown that the cross-sections for the pair production of scalar particles are about $10^{15}$ times larger than that of the  associated production under the same conditions.\\
\hspace*{0.5cm}In the $\gamma\gamma\rightarrow hh /\phi\phi$ collisions, due to the main contribution of scalar unparticle on the propagator in the s-channel, the cross - sections decrease as $\sqrt{s}$ increases while the cross-sections for the associated production increase as $\sqrt{s}$ increases. This is because the unparticle couplings in u,t channels give the main contribution on the $\gamma\gamma \rightarrow Uh/U\phi$ scattering process. However, the production cross-sections in $\gamma\gamma \text{    }\rightarrow hh/\phi\phi$ collisions are much larger than that of  $\gamma\gamma \rightarrow Uh/U\phi $ collisions (about $10^{5}$ times) under the same conditions.\\
\hspace*{0.5cm}In the $gg\rightarrow hh/\phi\phi$ collisions, the  cross-sections for the  pair production of the scalar particle decrease rapidly first and then increase as $\sqrt{s}$ increases, while the cross-sections for the associated production increase as $\sqrt{s}$ increases, which is similar to $\gamma\gamma \rightarrow Uh/U\phi $ process.
 Numerical evaluation has shown that the cross-sections of the associated production in the $gg$ collisions are much larger than that in the $\gamma \gamma$ collisions under the same conditions. This is because the scalar couplings with gluon are larger than that with the photon. \\
\hspace*{0.5cm} Finally, we emphasize that in this work we have considered only on a theoretical basis, other problems concerning 
  the scalar unparticle signals at LHC the readers can see in detail in Ref.\cite{alie} .\\

{\bf Acknowledgements}: The work is supported in part by the National Foundation for Science and Technology Development (NAFOSTED) of Vietnam under Grant No. 103.01-2016.44.\\

\newpage
\begin{figure}[!htbp] \label{fig:eehpdu}
\begin{center}
\includegraphics[width= 10 cm,height= 5 cm]{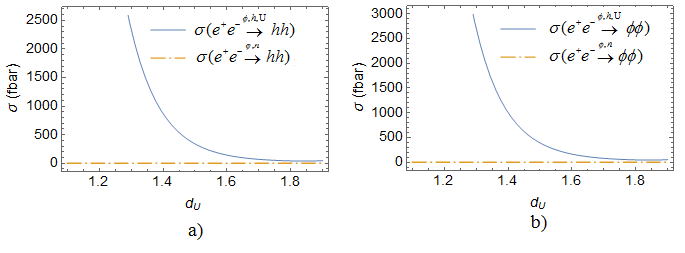}
\caption{The cross-section as a function of the $d_{U}$ in a) $e^{+}e^{-} \rightarrow hh$ collisions, (b) $e^{+}e^{-} \rightarrow \phi\phi$ collisions. The parameters are chosen as $\sqrt{s} = 500$ GeV, $\Lambda_{U} = 1000$ GeV.}
\end{center}
\end{figure}
\begin{figure}[!htbp] \label{Fig:eehp}
\begin{center}
\includegraphics[width= 12 cm,height= 10 cm]{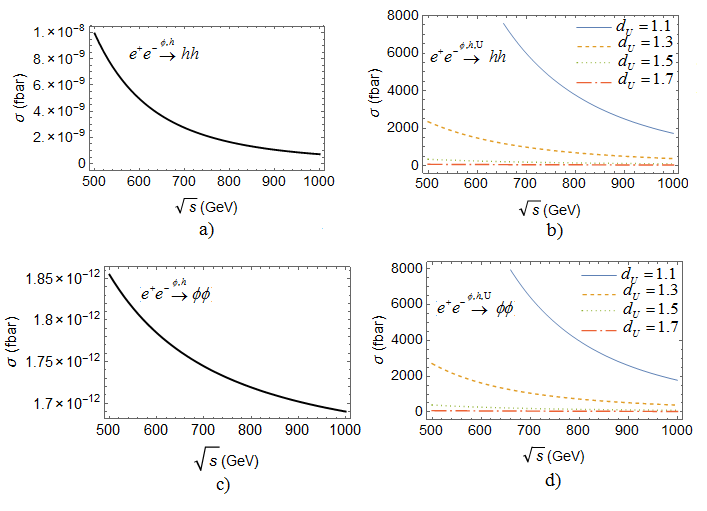}
\caption{The total cross-section as a function of the collision energy $\sqrt{s}$ in (a) $e^{+}e^{-} \rightarrow hh$ collision through $\phi, h$ propagators, (b) $e^{+}e^{-} \rightarrow hh$ collision through $\phi, h, U$ propagators, (c) $e^{+}e^{-} \rightarrow \phi\phi$ collision through $\phi, h$ propagators, (d) $e^{+}e^{-} \rightarrow \phi\phi$ collision through $\phi, h, U$ propagators. The energy scale $\Lambda_{U}$ is chosen as 1000 GeV. The various $d_{U}$ is chosen as 1.1, 1.3, 1.5, 1.7, respectively.}
\end{center}
\end{figure}
\begin{figure}[!htbp] \label{Fig:eehplu}
\begin{center}
\includegraphics[width= 10 cm,height= 5 cm]{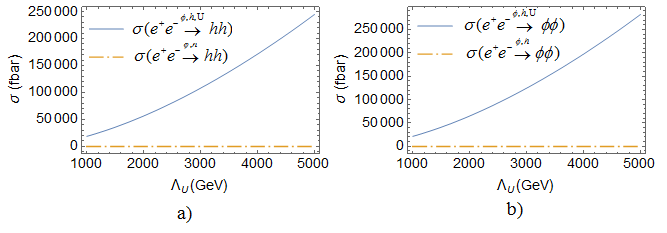}
\caption{The total cross-section as a function of the energy scale $\Lambda_{U}$ in (a) $e^{+}e^{-} \rightarrow hh$ collisions, (b) $e^{+}e^{-} \rightarrow \phi\phi$ collisions. The parameters are taken to be $\sqrt{s}$ = 500 GeV, $d_{U}$ = 1.1.}
\end{center}
\end{figure}
\begin{figure}[!htbp] \label{Fig:eedu}
\begin{center}
\includegraphics[width= 10 cm,height= 5 cm]{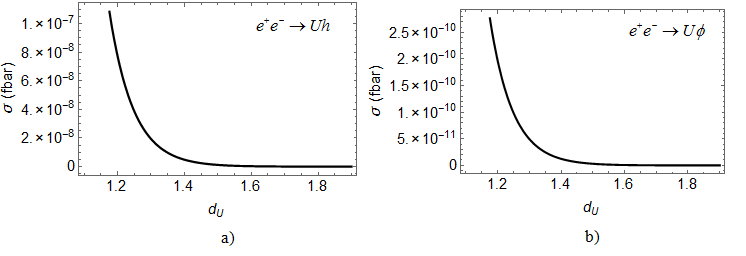}
\caption{The cross-section as a function of the $d_{U}$ in (a) $e^{+}e^{-} \rightarrow Uh$ collision, (b) $e^{+}e^{-} \rightarrow U\phi$ collision. The parameters are chosen as $\sqrt{s}$ = 500 GeV, $\Lambda_{U}$ = 1000 GeV.}
\end{center}
\end{figure}
\begin{figure}[!htbp] \label{Fig:ee}
\begin{center}
\includegraphics[width= 12 cm,height= 5 cm]{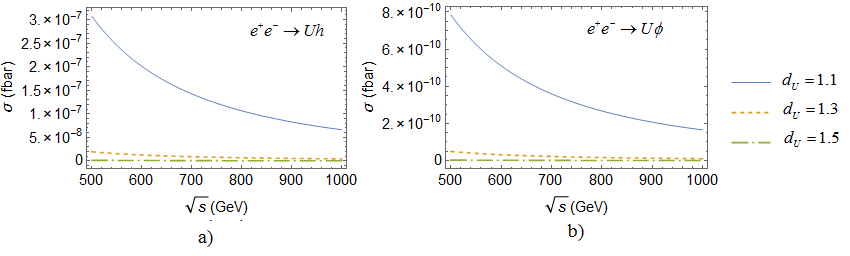}
\caption{The cross-section as a function of the collision energy $\sqrt{s}$ in (a) $e^{+}e^{-} \rightarrow Uh$ collision, (b) $e^{+}e^{-} \rightarrow U\phi$ collision. The energy scale $\Lambda_{U}$ is chosen as 1000 GeV. The $d_{U}$ is chosen as 1.1, 1.3, 1.5, respectively.}
\end{center}
\end{figure}
\begin{figure}[!htbp] \label{Fig:eelu}
\begin{center}
\includegraphics[width= 10 cm,height= 5 cm]{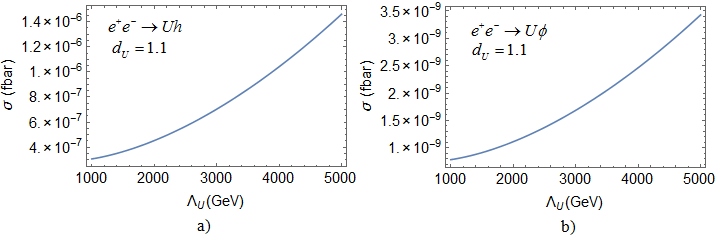}
\caption{The total cross-section as a function of $\Lambda_{U}$ with $d_{U} = 1.1$ in (a) $e^{+}e^{-} \rightarrow Uh$ collision, (b) $e^{+}e^{-} \rightarrow U\phi$ collision. The collision energy $\sqrt{s}$ is chosen as 500 GeV.}
\end{center}
\end{figure}
\begin{figure}[!htbp] \label{Fig:gmdu}
\begin{center}
\includegraphics[width= 10 cm,height= 5 cm]{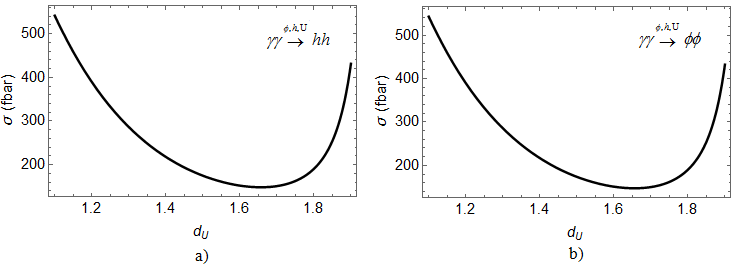}
\caption{The cross-section as a function of the $d_{U}$ in (a)$\gamma \gamma \rightarrow hh$ collision, (b)$\gamma \gamma \rightarrow \phi\phi$ collision. The parameters are chosen as $\sqrt{s}$ = 3000 GeV, $\Lambda_{U}$ = 1000 GeV.}
\end{center}
\end{figure}
\begin{figure}[!htbp] \label{Fig:gmcs}
\begin{center}
\includegraphics[width= 10 cm,height= 5 cm]{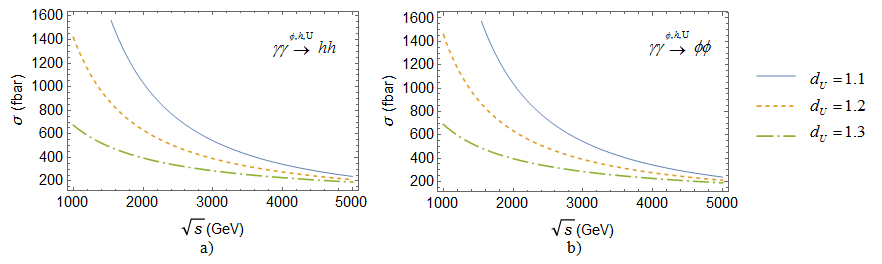}
\caption{The total cross-sections as a function of the collision energy $\sqrt{s}$ in (a) $\gamma \gamma \rightarrow hh$ collision through $\phi, h, U$ propagators, (b)$\gamma \gamma \rightarrow \phi\phi$ collision through $\phi, h, U$ propagators. The energy scale $\Lambda_{U}$ is chosen as 1000 GeV. The various $d_{U}$ is chosen as 1.1, 1.2, 1.3, respectively.}
\end{center}
\end{figure}
\begin{figure}[!htbp] \label{Fig:gmlu}
\begin{center}
\includegraphics[width= 10 cm,height= 5 cm]{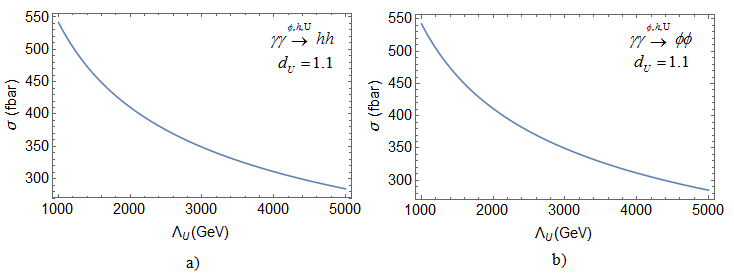}
\caption{The total cross-section as a function of $\Lambda_{U}$ with $d_{U}$ = 1.1 in (a)$\gamma \gamma \rightarrow hh$ collision, (b)$\gamma \gamma \rightarrow \phi\phi$ collision. The collision energy $\sqrt{s}$ is chosen as 3000 GeV.}
\end{center}
\end{figure}
\begin{figure}[!htbp] \label{Fig:gmudu}
\begin{center}
\includegraphics[width= 10 cm,height= 5 cm]{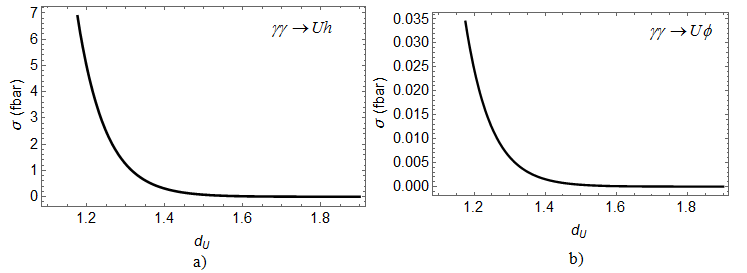}
\caption{The cross-section as a function of the $d_{U}$ in (a)$\gamma \gamma \rightarrow Uh$ collision, (b)$\gamma \gamma \rightarrow U\phi$ collision. The parameters are taken to be $\sqrt{s} = 3000$ GeV, $\Lambda_{U} = 1000$ GeV.}
\end{center}
\end{figure}
\begin{figure}[!htbp] \label{Fig:gmucs}
\begin{center}
\includegraphics[width= 10 cm,height= 5 cm]{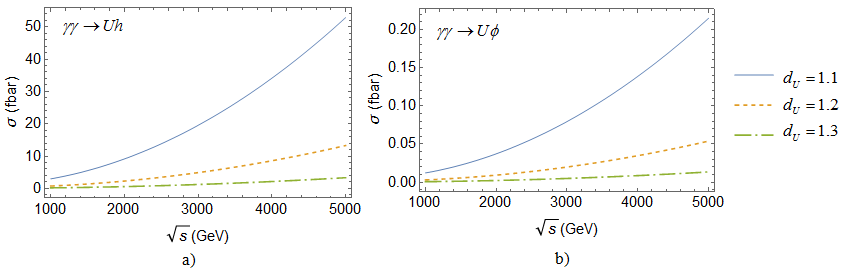}
\caption{The total cross-section as a function of the collision energy $\sqrt{s}$ in (a) $\gamma \gamma \rightarrow Uh$ collision, (b) $\gamma \gamma \rightarrow U\phi$ collision. The energy scale $\Lambda_{U}$ is chosen as 1000 GeV. The $d_{U}$ is chosen as 1.1, 1.2, 1.3, respectively.}
\end{center}
\end{figure}
\begin{figure}[!htbp] \label{Fig:gmulu}
\begin{center}
\includegraphics[width= 10 cm,height= 5 cm]{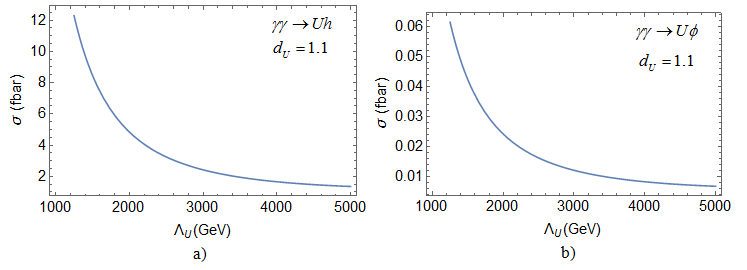}
\caption{The total cross-section as a function of $\Lambda_{U}$ with $d_{U}$ = 1.1 in (a) $\gamma \gamma \rightarrow Uh$ collision, (b) $\gamma \gamma \rightarrow U\phi$ collision . The collision energy $\sqrt{s}$ is chosen as 3000 GeV. }
\end{center}
\end{figure}
\begin{figure}[!htbp] \label{Fig:gdu}
\begin{center}
\includegraphics[width= 10 cm,height= 5 cm]{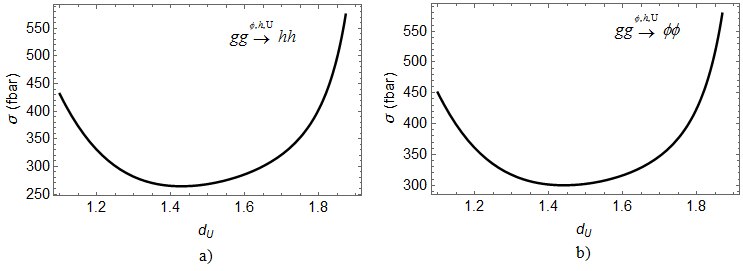}
\caption{The cross-section as a function of the $d_{U}$ in (a)$gg \rightarrow hh$ collision, (b)$gg \rightarrow \phi\phi$ collision. The parameters are taken to be $\sqrt{s} = 3000$ GeV, $\Lambda_{U} = 1000$ GeV.}
\end{center}
\end{figure}
\begin{figure}[!htbp] \label{Fig:gcs}
\begin{center}
\includegraphics[width= 10 cm,height= 5 cm]{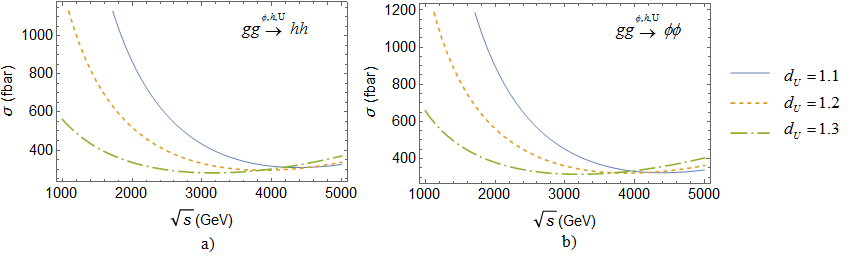}
\caption{The total cross-sections as a function of the collision energy $\sqrt{s}$ in (a) $gg \rightarrow hh$ collision through $\phi, h, U$ propagators, (b) $gg \rightarrow \phi\phi$ collision through $\phi, h, U$ propagators. The energy scale $\Lambda_{U}$ is chosen as 1000 GeV. The various $d_{U}$ is chosen as 1.1, 1.2, 1.3, respectively.}
\end{center}
\end{figure}
\begin{figure}[!htbp] \label{Fig:glu}
\begin{center}
\includegraphics[width= 10 cm,height= 5 cm]{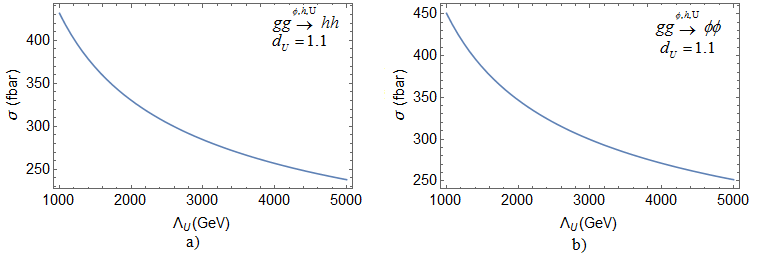}
\caption{The total cross-section as a function of $\Lambda_{U}$ in (a)$gg \rightarrow hh$ collision, (b)$gg \rightarrow \phi\phi$ collision. The parameters are chosen as  $\sqrt{s}$ = 3000 GeV, $d_{U}$ = 1.1.}
\end{center}
\end{figure}
\begin{figure}[!htbp] \label{Fig:gugu}
\begin{center}
\includegraphics[width= 10 cm,height= 4 cm]{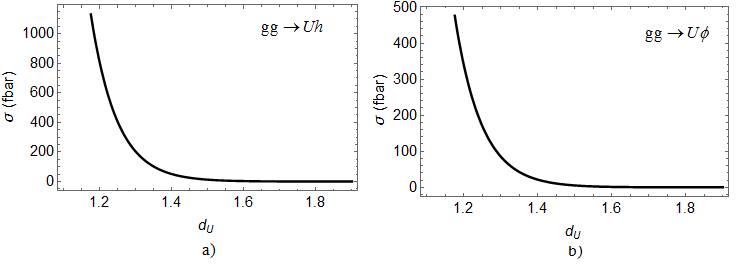}
\caption{The cross-section as a function of the $d_{U}$ in (a)$gg \rightarrow Uh$ collision, (b)$gg \rightarrow U\phi$ collision. The parameters are taken to be $\sqrt{s} = 3000$ GeV, $\Lambda_{U} = 1000$ GeV.}
\end{center}
\end{figure}
\begin{figure}[!htbp] \label{Fig:gucs}
\begin{center}
\includegraphics[width= 10 cm,height= 4 cm]{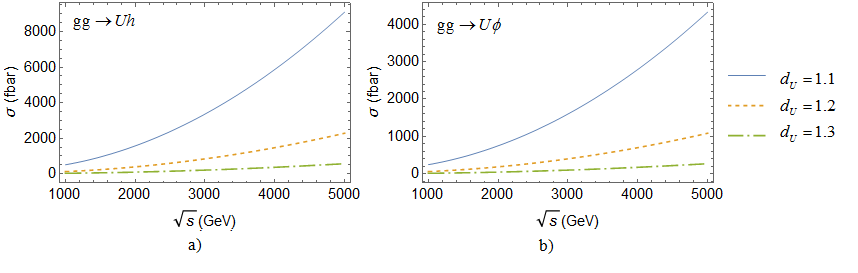}
\caption{The total cross-section as a function of the collision energy $\sqrt{s}$ in (a) $gg \rightarrow Uh$ collision, (b) $gg \rightarrow U\phi$ collision. The energy scale $\Lambda_{U}$ is chosen as 1000 GeV. The $d_{U}$ is chosen as 1.1, 1.2, 1.3, respectively.}
\end{center}
\end{figure}
\begin{figure}[!htbp] \label{Fig:gulu}
\begin{center}
\includegraphics[width= 10 cm,height= 4 cm]{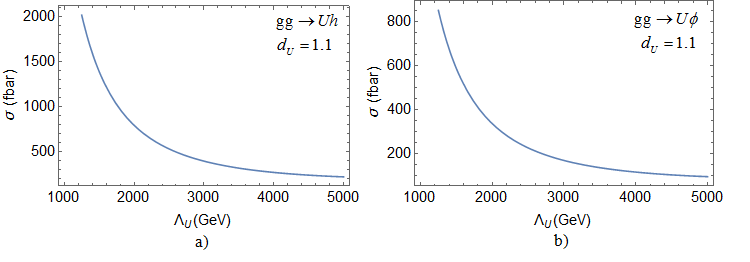}
\caption{The total cross-section as a function of $\Lambda_{U}$ in (a) $gg \rightarrow Uh$ collision, (b) $gg \rightarrow U\phi$ collision. The parameters are taken to be $\sqrt{s}$ = 3000 GeV, $d_{U}$ = 1.1. }
\end{center}
\end{figure}
\newpage
\begin{table}[!htbp] \label{Table:eecans11}
\begin{center}
\caption{Some typical values for the cross-section with the contribution of the scalar unparticle in the $e^{+}e^{-}$ collisions at the ILC.
 The parameters are chosen as $d_{U}$ = 1.1 and the masses $m_{h} = 125$ GeV, $m_{\phi} = 10$ GeV.}
\includegraphics[width= 15 cm,height= 5 cm]{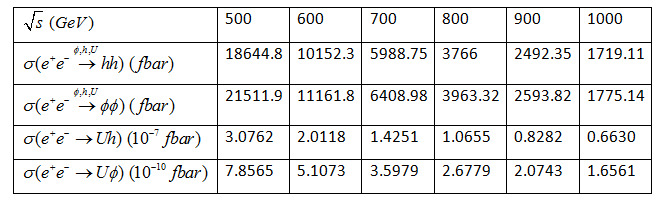}
\end{center}
\end{table}
\begin{table}[!htbp] \label{Table:gammacans11}
\begin{center}
\caption{Some typical values for the cross-section with the contribution of the scalar unparticle in the $\gamma\gamma$ collisions at the CLIC. The parameters are chosen as $d_{U}$ = 1.1 and the masses $m_{h} = 125$ GeV, $m_{\phi} = 10$ GeV.}
\includegraphics[width= 15 cm,height= 5 cm]{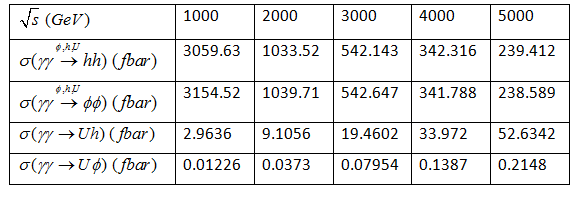}
\end{center}
\end{table}
\begin{table}[!htbp] \label{Table:gluoncans11}
\begin{center}
\caption{Some typical values for the cross-section with the contribution of the scalar unparticle in the $gg$ collisions at the CLIC.
 The parameters are chosen as $d_{U}$ = 1.1 and the masses $m_{h} = 125$ GeV, $m_{\phi} = 10$ GeV.}
\includegraphics[width= 15 cm,height= 5 cm]{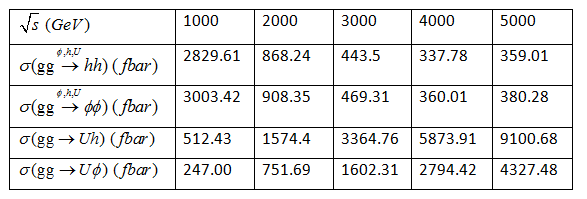}
\end{center}
\end{table}
\newpage

\begin{center} 
 { APPENDIX A: Feynman diagrams}\\
 \end{center}
\begin{figure}[!htbp] \label{Fig:fehp}
\begin{center}
\includegraphics[width= 14 cm,height= 4 cm]{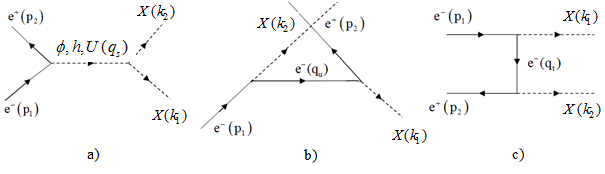}
\caption{Feynman diagrams for $e^{+}e^{-} \rightarrow hh/\phi\phi$ collisions through $\phi, h, U$ propagators, X stands for the Higgs or radion. The figures (a), (b), (c) representing the s, u, t-channel exchange, respectively.}
\end{center}
\end{figure}
\begin{figure}[!htbp] \label{Fig:feuhp}
\begin{center}
\includegraphics[width= 14 cm,height= 4 cm]{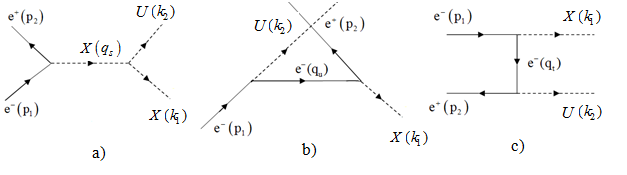}
\caption{Feynman diagrams for $e^{+}e^{-} \rightarrow Uh/U\phi$ collisions, X stands for the Higgs or radion. The figures (a), (b), (c) representing the s, u, t-channel exchange, respectively.}
\end{center}
\end{figure}
\begin{figure}[!htbp] \label{Fig:fgm}
\begin{center}
\includegraphics[width= 14 cm,height= 4 cm]{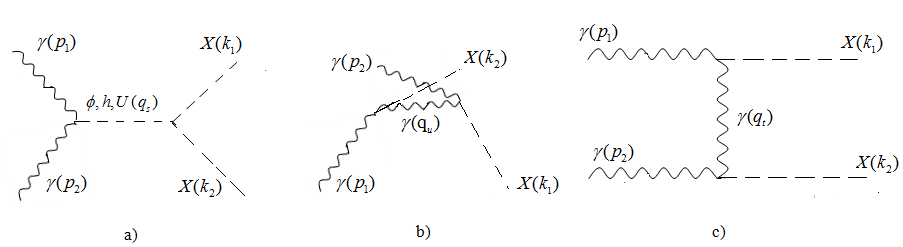}
\caption{Feynman diagrams for $\gamma\gamma \rightarrow \phi\phi/hh$ collisions through $\phi, h, U$ propagators, X stands for the Higgs or radion. The figures (a), (b), (c) representing the s, u, t-channel exchange, respectively.}
\end{center}
\end{figure}
\begin{figure}[!htbp] \label{Fig:fgmu}
\begin{center}
\includegraphics[width= 14 cm,height= 4 cm]{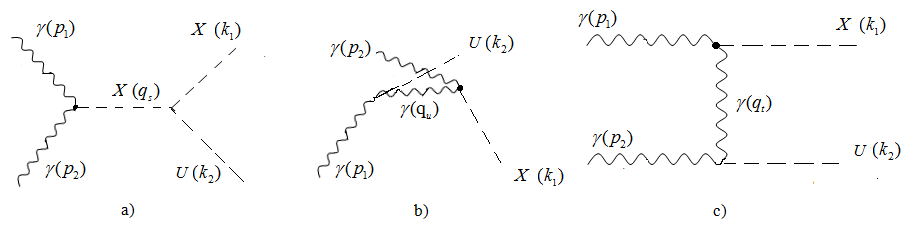}
\caption{Feynman diagrams for $\gamma\gamma \rightarrow Uh/U\phi$ collisions, X stands for the Higgs or radion. The figures (a), (b), (c) representing the s, u, t-channel exchange, respectively.}
\end{center}
\end{figure}
\begin{figure}[!htbp] \label{Fig:fg}
\begin{center}
\includegraphics[width= 14 cm,height= 4 cm]{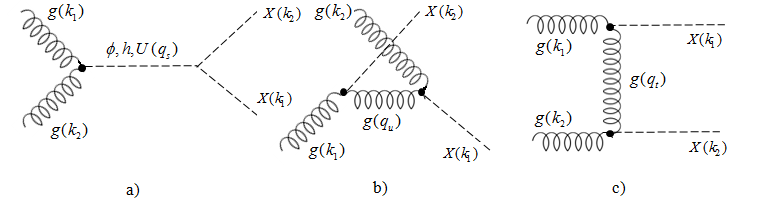}
\caption{Feynman diagrams for $gg \rightarrow \phi\phi/hh$ collisions through $\phi, h, U$ propagators, X stands for the Higgs or radion. The figures (a), (b), (c) representing the s, u, t-channel exchange, respectively.}
\end{center}
\end{figure}
\begin{figure}[!htbp] \label{Fig:fgu}
\begin{center}
\includegraphics[width= 14 cm,height= 4 cm]{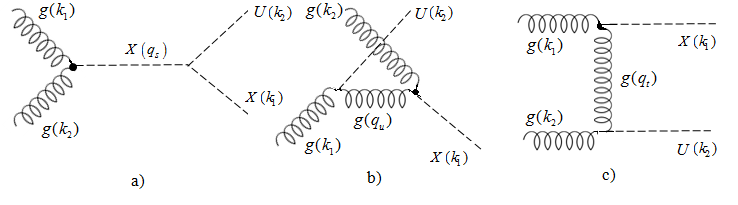}
\caption{Feynman diagrams for $gg \rightarrow Uh/U\phi$ collisions, X stands for the Higgs or radion. The figures (a), (b), (c) representing the s, u, t-channel exchange, respectively.}
\end{center}
\end{figure}

 \end{document}